\documentclass[preprint,prb]{revtex4}

\usepackage{graphicx}
\usepackage{dcolumn}
\usepackage{float}
\usepackage{amsmath}

\begin{document}

\title{\boldmath Bosonic spectrum of a correlated multiband system, BaFe$_{1.80}$Co$_{0.20}$As$_2$, obtained via infrared spectroscopy \unboldmath}

\author{Chandan Kumar Panda$^{1,+}$} \author{Hong Gu Lee$^{1,+}$} \author{Hwiwoo Park$^1$} \author{Soon-Gil Jung$^{2}$} \author{Ki-Young Choi$^{3}$} \author{Jungseek Hwang$^1$}
\email{jungseek@skku.edu}
\affiliation{$^1$Department of Physics, Sungkyunkwan University, Suwon, Gyeonggi-do 16419, Republic of Korea\\  $^2$Department of Physics Education, Sunchon National University, Suncheon 57922, Republic of Korea \\$^3$Defense Ai Technology Center, Agency for Defense Development, Daejeon 34186, Republic of Korea}
\date{\today}

\begin{abstract}

We investigated a single crystal BaFe$_{2-x}$Co$_x$As$_2$ (Co-doped BaFe$_2$As$_2$: Co-doped Ba122) with $x =$ 0.20 using infrared spectroscopy. We obtained the bosonic spectrum from the measured spectrum using an extended Drude-Lorentz model for the normal state and a two-parallel-channel approach for the superconducting (SC) state, based on the generalized Allen formula. The coupling constant, maximum SC transition temperature, SC coherence length, and upper critical field were extracted from the bosonic spectrum. The superfluid plasma frequency and the London penetration depth were obtained from the optical conductivity. We compared the physical quantities of Co-doped Ba122 and K-doped Ba122 and found some interesting differences. Our results may be helpful for understanding superconductivity in doped Ba122 systems and may provide useful information on doped Ba122 systems for their applications.
\\

\noindent {\bf Keywords:} Fe-pnictides, Co-doped Ba122, bosonic spectrum, optical conductivity, extended Drude-Lorentz model, two-parallel-channel approach
\\

\noindent *Correspondence author \\
  Email: jungseek@skku.edu (Jungseek Hwang)

\end{abstract}

%\pacs{74.25.-q, 74.25.Gz, 74.25.Kc}

\maketitle

\section*{Introduction}

Decades have passed since the discovery of novel Fe-based high-temperature superconductors \cite{kamihara:2006,kamihara:2008}. Numerous experimental and theoretical investigations have been performed since their discovery  \cite{ding:2008,qazilbash:2009,yang:2009a,mazin:2010a,basov:2011,dai:2013,nakajima:2014,bang:2017}. Nonetheless, the microscopic pairing mechanism still remains elusive.  The Fe-based superconductors have been known to be correlated multiband materials\cite{subedi:2008,ding:2008,qazilbash:2009}. Measured infrared spectrum can carry correlation information that appears as a band renormalization. The correlation between electrons can be described by interacting electrons exchanging force-mediating bosons. The correlation information (or bosonic spectrum) of Fe-based superconductors has been extracted from measured optical scattering rates by using the extended Drude model and either the Eliashberg or Allen approaches  \cite{qazilbash:2009,yang:2009a,wu:2010,charnukha:2011,moon:2012,hwang:2015}.
Benfatto {\it et al.} suggested that low-energy interband transitions in the infrared region should be included in the analysis \cite{benfatto:2011}. A reverse process \cite{hwang:2015a} and either the extended Drude-Lorentz model\cite{lee:2022} or two-parallel-channel approach \cite{hwang:2016,lee:2022a} have been proposed to extract the bosonic spectrum from measured optical conductivities. The extracted bosonic spectrum can provide information on various important physical quantities, such as the coupling constant between charge carriers, the maximum superconducting (SC) transition temperature, and the SC coherent length \cite{hwang:2021}. The asymmetric phase diagram of BaFe$_2$As$_2$ (Ba122) is quite similar to that of cuprates \cite{basov:2011}; this indicates that the two material systems may share a microscopic pairing mechanism. The hole-doped (K-doped) Ba122 exhibits a higher superconducting (SC) transition temperature and a broader SC dome than its electron-doped (Co-doped or Ni-doped) counterpart \cite{neupane:2011}. In previous literature \cite{neupane:2011}, the authors have demonstrated that the asymmetric phase diagram can be attributed to the screening effect and chemical potential shift caused by doping. However, the study did not explain why the $T_c$s are different between hole- and electron-doped Ba122 systems.

In this study, we investigated and compared a Co-doped (electron-doped) Ba122 (BaFe$_{1.80}$Co$_{0.20}$As$_2$) with a K-doped (hole-doped) Ba122 (Ba$_{0.49}$K$_{0.51}$Fe$_2$As$_2$) using infrared spectroscopy. Both systems were similarly overdoped. The K-doped Ba122 system has been studied using similar approaches, and the results have been reported \cite{lee:2022,lee:2022a}. The superfluid plasma frequency and the London penetration depth were obtained from the measured optical conductivity of the Co-doped Ba122. The bosonic spectrum was obtained from the optical conductivity of the Co-doped Ba122 using the reverse process \cite{hwang:2015a} and the extended Drude-Lorentz model\cite{lee:2022} for the normal state, and using the reverse process and the two-parallel-channel approach \cite{hwang:2016} for the SC state. In the extraction process of the bosonic spectra of the Co-doped Ba122, we had to include a large impurity scattering rate (10 meV) and two Lorentz modes at a low frequency region below 400 cm$^{-1}$, which were absent in similar analyses for K-doped Ba122 and LiFeAs \cite{hwang:2016,lee:2022,lee:2022a}. To the best of our knowledge, this study is the first to apply the extended Drude-Lornetz model analysis and the two-parallel band approach to a Co-doped Ba122 system. The coupling constant ($\lambda$), the maximum SC transition temperature ($T_c^{Max}$), the SC coherence length ($\xi_{SC}$), and the upper critical filed were obtained from the extracted bosonic spectrum. These quantities are important for applications of the material system. The obtained $T_c^{Max}$ was found to be larger than the actual $T_c$ measured by the DC transport technique. We compared the physical quantities of the Co-doped Ba122 with those of K-doped Ba122. Both the coupling constant and the SC coherence length of Co-doped Ba122 were larger than those of the K-doped one because the bosonic spectrum of the Co-doped Ba122 is narrowly spread from zero energy compared with the K-doped one. The $T_c^{Max}$ of Co-doped Ba122 was found to be smaller than that of K-doped Ba122, which is also closely related to the energy-dependent distribution of the bosonic spectrum. We believe that our results may provide helpful insights into the asymmetric phase diagram of Ba122 systems, the microscopic pairing mechanism of the Cooper pairs, and information for their applications.

\section*{Experiment}

A high-quality single crystal of BaFe$_{1.80}$Co$_{0.20}$As$_2$ was investigated using infrared spectroscopy. The single crystal sample was grown using a high-temperature self-flux method. The detailed description of the growth method can be found in previous studies \cite{canfield:1992,wang:2009,ni:2009}. The measured x-ray diffraction pattern of the crystal sample is shown in the Supplementary Materials (Fig. S1). The sample has an area of roughly 2$\times$2 mm$^2$ with a thickness $\leq$ 200 $\mu$m. A thin brass plate was placed between the sample and a sample cone to prevent bending of the thin sample caused by an epoxy contraction when it was cooled down. The SC transition temperature was determined to be 22.0 K using a DC transport (4-probe) measurement (see the upper right inset of Fig. \ref{fig1}, and see also Fig. S2 in the Supplementary Materials). It is worth noting that there is a kink near 22.5 K in the resistivity curve, which may indicate the possible existence of two SC phases. Because the difference in $T_c$ between the two SC phases is small ($\sim$ 0.5 K), the two SC phases are very close to each other. An {\it in-situ} gold evaporation method \cite{homes:1993} and a continuous liquid Helium flow optical cryostat were used to obtain accurate reflectance spectra of the sample in a wide spectral range (50 - 8000 cm$^{-1}$) at various temperatures. A commercial Fourier-transform infrared (FTIR)-type spectrometer (Vertex 80v, Bruker) was used to measure the reflectance spectra. The Kramers-Kronig relation, the Fresnel formula for normal incidence, and relations between optical constants such as the index of refraction, dielectric function, and the optical conductivity \cite{wooten,tanner:2019} were used to obtain the optical conductivity from the measured reflectance. In order to perform the Kramers-Kronig analysis, the measured reflectance was extrapolated to both zero and infinity. For the extrapolation to zero, we used the Hagen-Rubens relation, i.e., $1-R(\omega) \propto \sqrt{\omega}$ for the normal state, and $1-R(\omega) \propto \omega^4$ for the superconducting state. For the extrapolation to infinity, the measured reflectance was extended up to 53,000 cm$^{-1}$ using previously published data \cite{tu:2010}. Above 53,000 cm$^{-1}$ to 10$^6$ cm$^{-1}$, the extrapolation was conducted using $R(\omega) \propto \omega^{-1}$. Above 10$^6$ cm$^{-1}$, the free electron behavior, i.e., $R(\omega) \propto \omega^{-4}$, was assumed.

\section*{Results and discussion}

\subsection*{Reflectance and optical conductivity}

\begin{figure}[!htbp]
  \vspace*{-1.0 cm}%
 \centerline{\includegraphics[width=7.0 in]{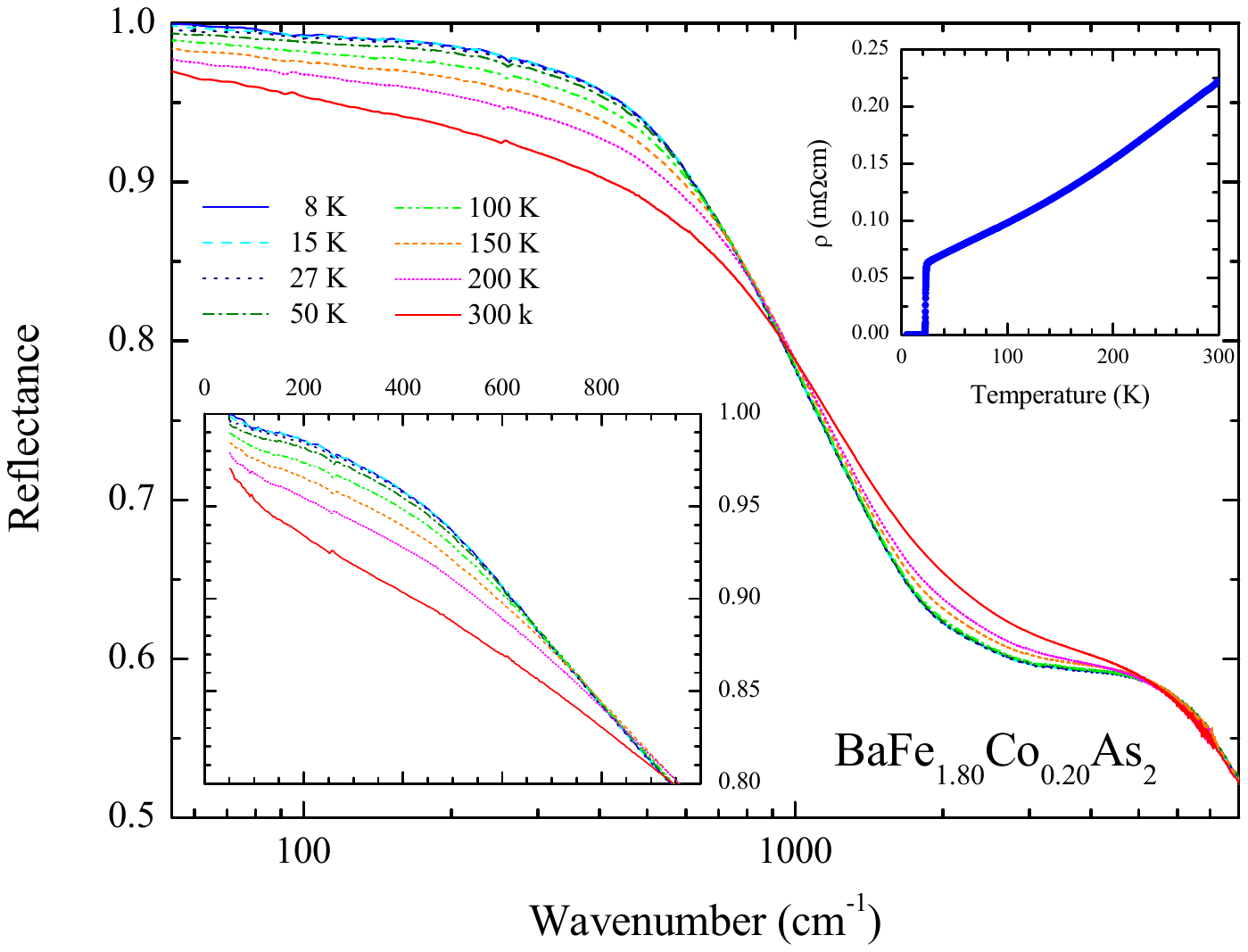}}%
  \vspace*{-1.0 cm}%
\caption{(Color online) Measured {\it ab} plane reflectance spectra of single crystal BaFe$_{1.80}$Co$_{0.20}$As$_2$ at various temperatures in a wide spectral range from 50 cm$^{-1}$ to 8000 cm$^{-1}$. In the lower left inset, a magnified view of the reflectance in a low frequency region below 1000 cm$^{-1}$ is shown. In the upper right inset, the measured DC resistivity as a function of temperature is shown.}
\label{fig1}
\end{figure}

Figure \ref{fig1} shows the measured $ab$ plane reflectance of BaFe$_{1.80}$Co$_{0.20}$As$_2$ at various temperatures in a temperature range of 8 - 300 K. The overall temperature-dependent trend of the reflectance of our over-Co-doped Ba122 is similar to the reported ones of Co-doped Ba122 systems \cite{tu:2010,kim:2010,lobo:2010,heumen:2010}, which were underdoped or optimally doped. However, the overall level of reflectance of our over-Co-doped Ba122 is higher than the reported ones. As we expected, the reflectance shows a metallic behavior, i.e., the reflectance is enhanced at low frequencies below $\sim$ 800 cm$^{-1}$ when the temperature decreases. At 8 K, the reflectance exhibited a sharp increase below $\sim$ 80 cm$^{-1}$, which is a signature for the formation of the SC gap (see Fig. S3 in the Supplementary Materials for a better view of the feature).

\begin{figure}[!htbp]
  \vspace*{-1.0 cm}%
 \centerline{\includegraphics[width=7.0 in]{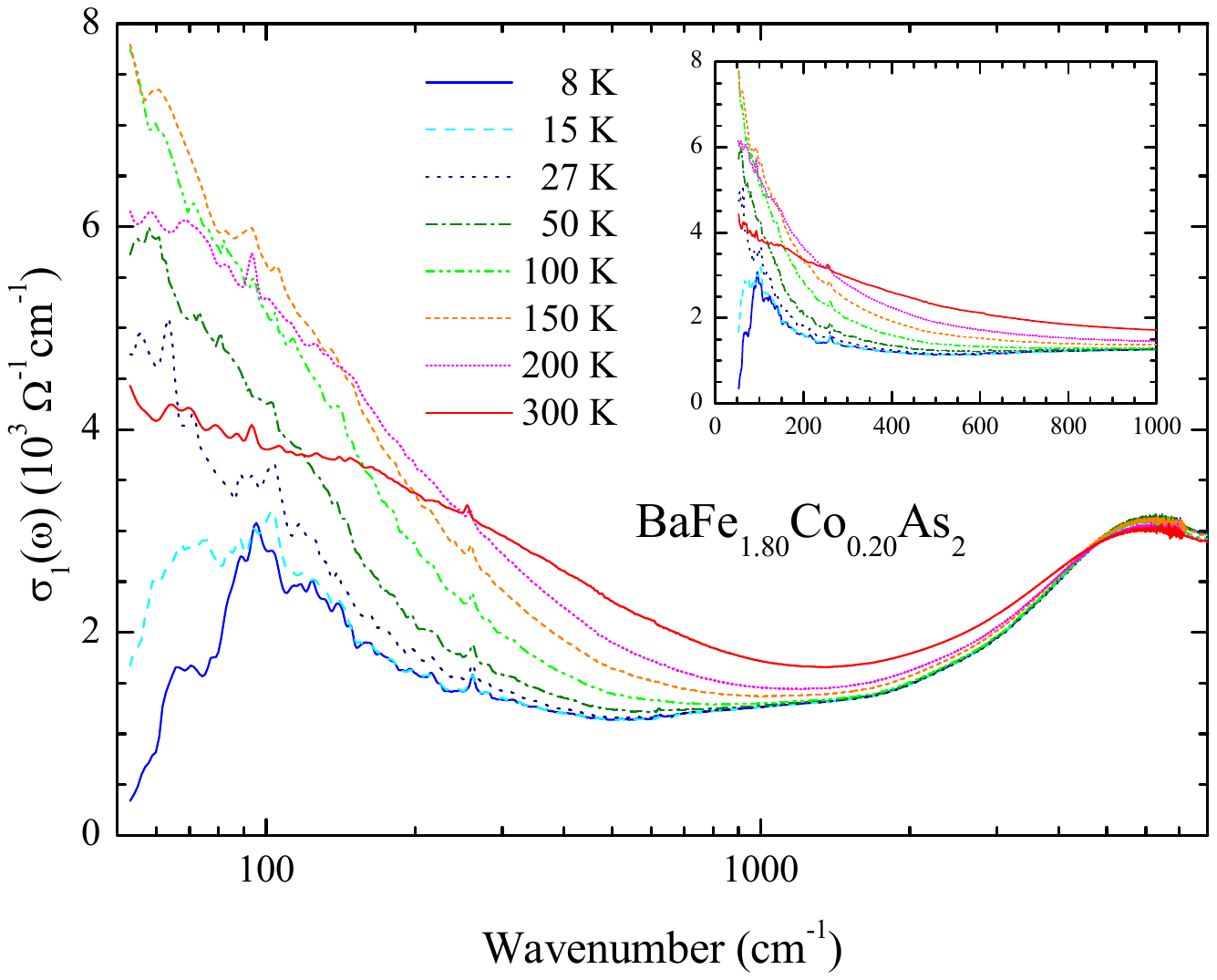}}%
  \vspace*{-1.0 cm}%
\caption{(Color online) Optical conductivity of BaFe$_{1.80}$Co$_{0.20}$As$_2$ at various temperatures. The inset illustrates a magnified view of the conductivity below 1000 cm$^{-1}$.}
\label{fig2}
\end{figure}

Fig. \ref{fig2} shows the optical conductivity obtained from the measured reflectance using the Kramers-Kronig analysis. As shown in the figure, the temperature-dependent trends of typical and correlated metals in the low frequency region are not monotonic below 200 cm$^{-1}$ because the thermally excited phonons increase the scattering rate and the total spectral weight is conserved. This results in the temperature-dependent spectral weight redistribution. For the SC state, the superfluid spectral weight appears at $\omega =$ 0 because the scattering rate of the electrons (or spectral weight) involved in the superfluid is absolutely zero. The spectral weight does not appear in the finite frequency region and is called the "missing spectral weight", from which the superfluid spectral weight (or plasma frequency) can be estimated. The optical conductivities of the SC state (at 8 K and 15 K) show spectral weight suppressions below the SC gap, i.e., $\sim$ 50 cm$^{-1}$. In the inset in Fig. \ref{fig2}, a magnified view of the conductivity below 1000 cm$^{-1}$ is shown. In addition, a well-known infrared active phonon \cite{akrap:2009} can be seen near 260 cm$^{-1}$.

\begin{figure}[!htbp]
  \vspace*{-1.0 cm}%
 \centerline{\includegraphics[width=7.0 in]{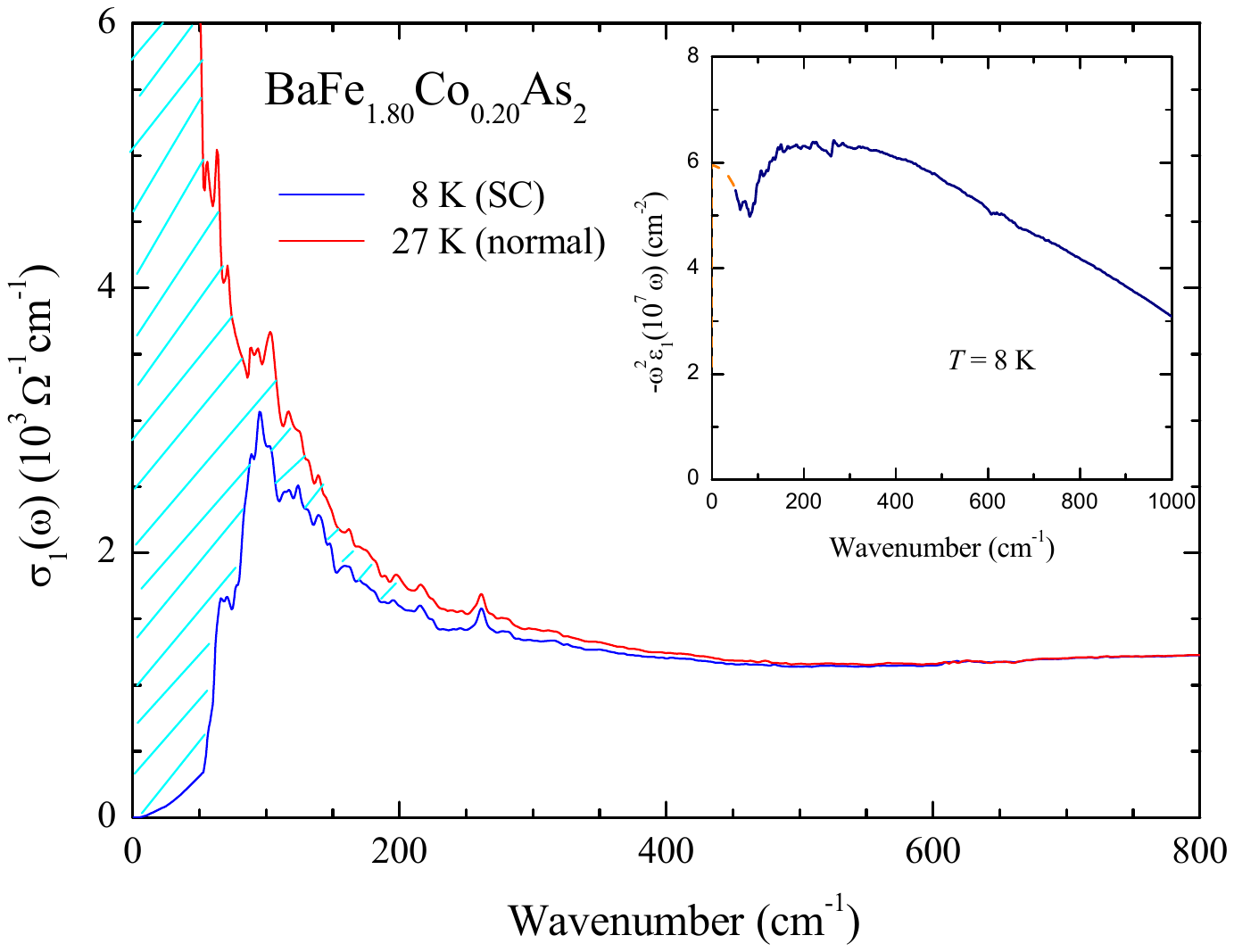}}%
  \vspace*{-1.0 cm}%
\caption{(Color online) Optical conductivity spectra of BaFe$_{1.80}$Co$_{0.20}$As$_2$ at two temperatures: 8 K (SC state) and 27 K (normal state). In the insets, $-\omega^2 \varepsilon_1(\omega)$ as a function of $\omega$ is shown.}
\label{fig3}
\end{figure}

The superfluid plasma frequency ($\omega_{sp}$) of 8 K was obtained by two independent methods: one using the real part of optical conductivity ($\sigma_1(\omega)$), known as the Ferrell-Glover-Tinkham (FGT) sum rule \cite{glover:1956,ferrell:1958,tinkham:1975}, and the other using the imaginary part of optical conductivity ($\sigma_2(\omega)$) \cite{lee:2022a}. In the FGT sum rule, the superfluid plasma frequency is described as $\omega_{sp} = \sqrt{(120/\pi)\int_{0^+}^{\infty}[\sigma_{1, N}(\omega)-\sigma_{1, SC}(\omega)] d\omega}$, where $\sigma_{1,N}(\omega)$ and $\sigma_{1, SC}(\omega)$ denote the real parts of optical conductivity of the normal and SC states, respectively. As shown in Fig. \ref{fig3}, the integration is equal to the hatched area in Fig. \ref{fig3}. In the second method, $\omega_{sp} = \lim_{\omega \rightarrow 0} \sqrt{[-\omega^2\varepsilon_1(\omega)]}$, where $\varepsilon_1(\omega)$ denotes the dielectric function, which is related to $\sigma_2(\omega)$ as $\varepsilon_1(\omega) = (60/\omega)\sigma_2(\omega)$. In the inset of Fig. \ref{fig3}, $-\omega^2 \varepsilon_1(\omega)$ is displayed as a function of frequency. Note that the unit for all frequencies is cm$^{-1}$. By using the first and second methods, the superfluid plasma frequencies obtained are 7470 cm$^{-1}$ and 7710 cm$^{-1}$, respectively. The resulting $\omega_{sp}$ obtained by the two methods agree well with each other, even though the second method gives a slightly larger value than the first one. The London penetration depth ($\lambda_L$) can be obtained from the superfluid plasma frequency using the relation $\lambda = 1/(2\pi \omega_{sp})$. The London penetration depth obtained using the FGT sum rule was 213.2 nm, while that obtained using the other method was 206.5 nm; these values are comparable to the previous results of Ba122 compounds \cite{li:2008,tu:2010,lobo:2010,heumen:2010,kim:2010,yoon:2017}.

\subsection*{Bosonic spectrum}

The bosonic spectra were obtained from the measured optical conductivities at 27 K (normal state) using the extended Drude-Lorentz model \cite{lee:2022} and 8 K (SC state) using the two-parallel-channel approach \cite{hwang:2016}. In the extended Drude-Lorentz model, the complex optical conductivity ($\tilde{\sigma}(\omega)$) is described as follows \cite{lee:2022}:
\begin{equation}
  \tilde{\sigma}(\omega) = \tilde{\sigma}_{\mathrm{ED}}(\omega)+\tilde{\sigma}_{\mathrm{L_i}}(\omega),
\end{equation}
where $\tilde{\sigma}_{\mathrm{ED}}(\omega)$ denotes the extended Drude (ED) mode, which is obtained using the reverse process \cite{hwang:2015a}, and $\tilde{\sigma}_{\mathrm{L_i}}(\omega)$ denotes the {\it i}th component of the Lorentz modes, which can be used to describe the interband transitions or phonon modes \cite{wooten,tanner:2019}. In the two-parallel-channel approach, the complex optical conductivity ($\tilde{\sigma}(\omega)$) is described as follows \cite{hwang:2016}:
\begin{equation}
  \tilde{\sigma}(\omega) = \tilde{\sigma}_{\mathrm{Ch1}}(\omega)+\tilde{\sigma}_{\mathrm{Ch2}}(\omega)+\tilde{\sigma}_{L_i}(\omega),
\end{equation}
where $\tilde{\sigma}_{\mathrm{Ch1}}(\omega)$ and $\tilde{\sigma}_{\mathrm{Ch2}}(\omega)$ denote the complex optical conductivities of the channel 1 and channel 2, respectively. Note that each channel is a SC channel. $\tilde{\sigma}_{\mathrm{Ch1}}(\omega)$ and $\tilde{\sigma}_{\mathrm{Ch2}}(\omega)$ were obtained the reverse process \cite{hwang:2015a}, in which the optical scattering rate, $1/\tau^{op}(\omega)$, (or the imaginary part of the optical self-energy, $-2\Sigma^{op}_2(\omega)$) were obtained from an input bosonic spectrum using the generalized Allen's formulas, the real part of the optical self-energy ($-2\Sigma^{op}_1(\omega)$) was obtained from the imaginary one using the Kramers-Kronig relation between the real and imaginary parts of the optical self-energy, and eventually, the complex optical conductivity was obtained using the ED model formalism.

The imaginary part of the optical self-energy for the ED mode ($-2\Sigma^{op}_{\mathrm{ED},2}(\omega)$) is written as follows\cite{lee:2022}:
\begin{eqnarray}
-2\Sigma^{op}_{\mathrm{ED},2}(\omega) &=& \frac{\pi}{\omega} \int_0^{\infty}d\Omega \: I^2B(\Omega)\Big{[}2\omega \coth\Big{(}\frac{\Omega}{2T}\Big{)}-(\omega+\Omega)\coth\Big{(}\frac{\omega+\Omega}{2T}\Big{)} \nonumber \\ &+&(\omega-\Omega)\coth\Big{(}\frac{\omega-\Omega}{2T}\Big{)} + \frac{1}{\tau^{op}_{\mathrm{imp}}} \Big{]},
\end{eqnarray}
where $I^2B(\Omega)$ denotes the bosonic spectrum. Here $I$ denotes the coupling constant between the electron and a force mediated boson. $1/\tau^{op}_{\mathrm{imp}}$ denotes the impurity scattering rate. The imaginary part of the optical self-energy for each SC channel ($-2\Sigma^{op}_{\mathrm{Ch_i},2}(\omega)$) is written as follows\cite{hwang:2016}:
\begin{equation}
-2\Sigma^{op}_{\mathrm{Ch_i},2}(\omega) = \int^{\infty}_{0}\!\!d\Omega \:I^2B(\Omega) \:K(\omega,\Omega) + \frac{1}{\tau^{op}_{\mathrm{imp}}(\omega)},
\end{equation}
where $i$ denotes the channel number, which is either 1 or 2. $K(\omega,\Omega)$ denotes the kernel, which is written as follows \cite{allen:1971}:
\begin{eqnarray}
K(\omega,\Omega) &=& \frac{2\pi}{\omega}(\omega-\Omega)\Theta(\omega-2\Delta_{\mathrm{Ch_i},0}- \Omega) \nonumber \\ \!\!&\times& \!\!\! E\Big{(}\frac{\sqrt{(\omega-\Omega)^2\!-\!(2\Delta_{\mathrm{Ch_i},0})^2}}{\omega\!-\!\Omega}\Big{)},
\end{eqnarray}
where $\Theta(\omega)$ represents the Heaviside step function (i.e., 1 for $\omega \geq 0$ and 0 for $\omega < 0$), and $E(x)$ represents the complete elliptic integral of the second kind, where $x$ is dimensionless. $\Delta_{\mathrm{Ch_i},0}$ denotes the SC gap for the channel $i$. $1/\tau^{op}_{\mathrm{imp}}(\omega)$ denotes the impurity scattering rate for the $s$-wave SC state, which is frequency-dependent and is written as follows\cite{allen:1971}:
\begin{eqnarray}
\frac{1}{\tau^{op}_{imp}(\omega)} &=&\!\! \frac{1}{\tau^{op}_{\mathrm{imp}}} E\Big{(}\frac{\sqrt{\omega^2\!-\!(2\Delta_{\mathrm{Ch_i},0})^2}}{\omega}\Big{)}.
\end{eqnarray}
Note that the same $I^2B(\omega)$ was used for the two channels as in the previous studies \cite{hwang:2016,lee:2022a}.

Once the imaginary part of the optical self-energy is obtained, the corresponding real part can be obtained using the Kramers-Kronig relation between them, expressed as follows \cite{hwang:2016}:
\begin{equation}
-2\Sigma^{op}_1(\omega) = -\frac{2}{\pi} P\int^{\infty}_{0}\frac{\Omega[-2\Sigma^{op}_2(\Omega)], }{\Omega^2-\omega^2}d\Omega
\end{equation}
where $P$ denotes for the principal part of the improper integral. Here, $-2\Sigma^{op}_2(\Omega)$ can be either $-2\Sigma^{op}_{\mathrm{ED},2}(\omega)$ or $-2\Sigma^{op}_{\mathrm{Ch_i},2}(\omega)$, depending upon the electronic (either normal or SC) state. Eventually, the complex optical conductivity ($\tilde{\sigma}(\omega) \equiv \sigma_1(\omega) + i\sigma_2(\omega)$) was obtained for the extended Dude mode and the two SC channels using the ED model formalism as follows \cite{gotze:1972,allen:1977,puchkov:1996,hwang:2004}:
\begin{equation}
\tilde{\sigma}(\omega)=i\frac{\Omega_p^2}{4\pi}\frac{1}{\omega+[-2\tilde{\Sigma}^{op}(\omega)]},
\end{equation}
where $\Omega_p$ denotes the plasma frequency of charge carriers, which can be different for two different channels, and $\tilde{\Sigma}^{op}(\omega)$ $(\equiv \Sigma^{op}_1(\omega) + i\Sigma^{op}_2(\omega))$ denotes the complex optical self-energy. Here, $-2\tilde{\Sigma}^{op}(\Omega)$ can be either $-2\tilde{\Sigma}^{op}_{\mathrm{ED}}(\omega)$ or $-2\tilde{\Sigma}^{op}_{\mathrm{Ch_i}}(\omega)$, depending on the electronic (either normal or SC) state.

\begin{figure}[!htbp]
  \vspace*{-1.0 cm}%
 \centerline{\includegraphics[width=7.0 in]{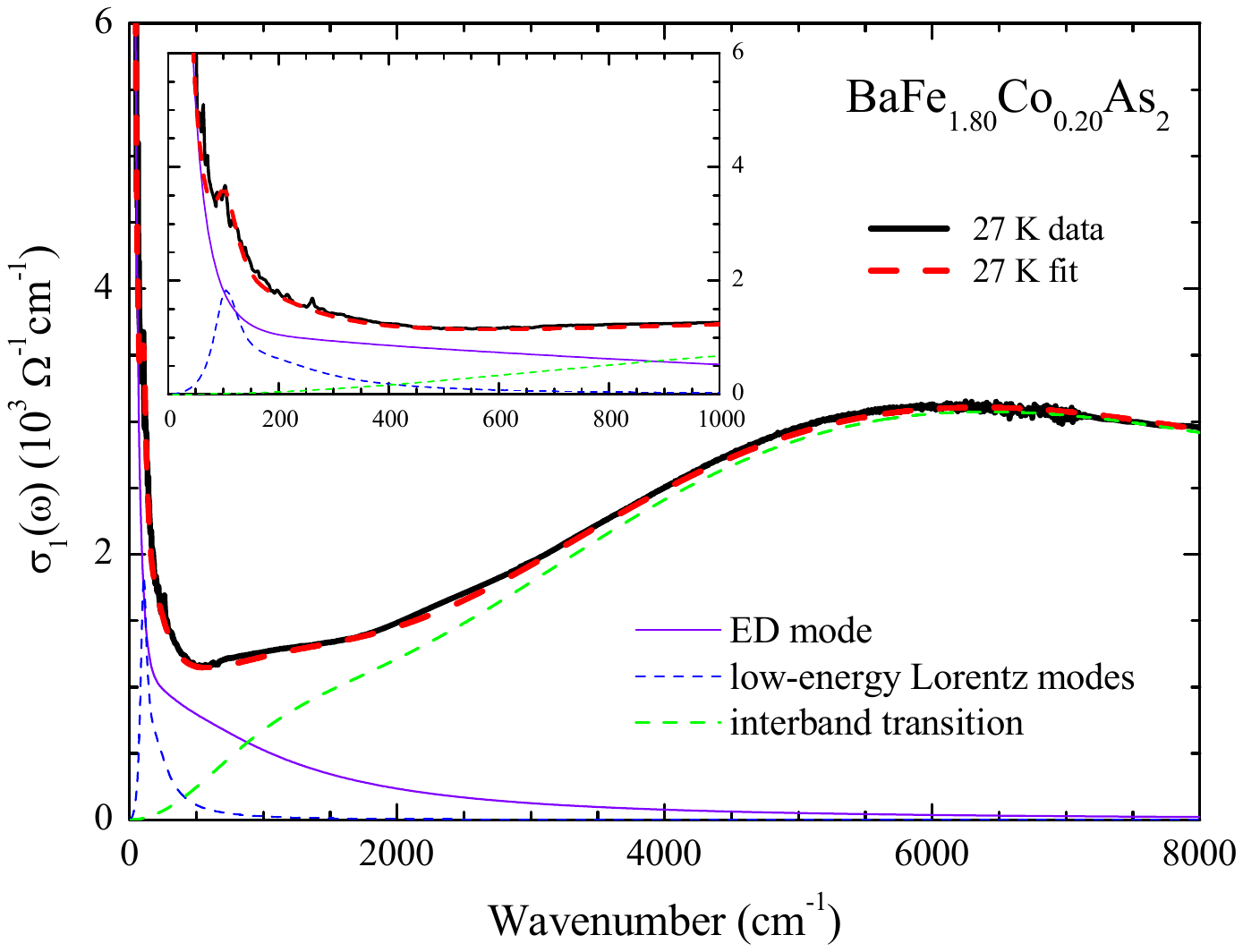}}%
  \vspace*{-1.0 cm}%
\caption{(Color online) Optical conductivity spectrum of BaFe$_{1.80}$Co$_{0.20}$As$_2$ at 27 K (normal state) and its fit in a wide spectral range up to 8000 cm$^{-1}$. In the inset, a magnified view of conductivity is shown to display the ED mode in the purple solid line and the two low-energy Lorentz modes in the blue dashed line.}
\label{fig4}
\end{figure}

The model $I^2B(\omega)$ was used in this study; it consists of two Gaussian peaks: one is sharp and located at a low frequency, and the other is broad and located at a high frequency. The same model bosonic spectrum has been used in previous studies as well \cite{lee:2022,lee:2022a}. Here, the sharp Gaussian is an optical mode, which may be associated with the magnetic resonance mode observed by the inelastic neutron scattering \cite{inosov:2010}. Fig. \ref{fig4} shows data and fit at 27 K (normal state) using the extended Drude-Lorentz model described above in a wide spectral range up to 8000 cm$^{-1}$. The resulting bosonic spectrum ($I^2B(\omega)$) is shown in Fig. \ref{fig6} (see the red dashed line). An extended Drude mode, two Lorentz modes, and an interband transition were needed for the fitting. Note that the interband transition consists of three Loremtz modes. The plasma frequency ($\Omega_{\mathrm{ED},p}$) and impurity scattering rate ($1/\tau^{op}_{\mathrm{imp}}$) for the ED mode were found to be 1.23 eV and 15 meV, respectively. The the inset of Fig. \ref{fig4} displays a magnified view of the ED mode and the two Lorentz modes in the blue dashed line located at low frequencies below 1000 cm$^{-1}$. Interestingly, two out of the five Lorentz modes are located in a very low frequency region below $\sim$400 cm$^{-1}$; these two modes were named low-energy Lorentz modes, as shown in the inset. Nevertheless, these low-energy Lorentz modes are absent in the K-doped Ba122 systems \cite{dai:2013,lee:2022,lee:2022a} (see also Fig. S4 and S5 in the Supplementary Materials) but have been observed in a similar frequency region by previous optical studies of Co-doped Ba122 systems \cite{lobo:2010,heumen:2010}. The origin of low-energy Lorentz modes is not yet clarified. Lobo {\it et al.} speculated that these Lorentz modes are the response of localized carriers induced by disorders \cite{lobo:2010}. The disorder is related to the FeAs plane lattice distortion caused by Co-doping in the Co-doped Ba122 systems because these low-energy Lorentz modes are absent in the K-doped Ba122 systems \cite{dai:2013,lee:2022,lee:2022a}, where the FeAs plane is intact by K-doping.

\begin{figure}[!htbp]
  \vspace*{-1.0 cm}%
 \centerline{\includegraphics[width=7.0 in]{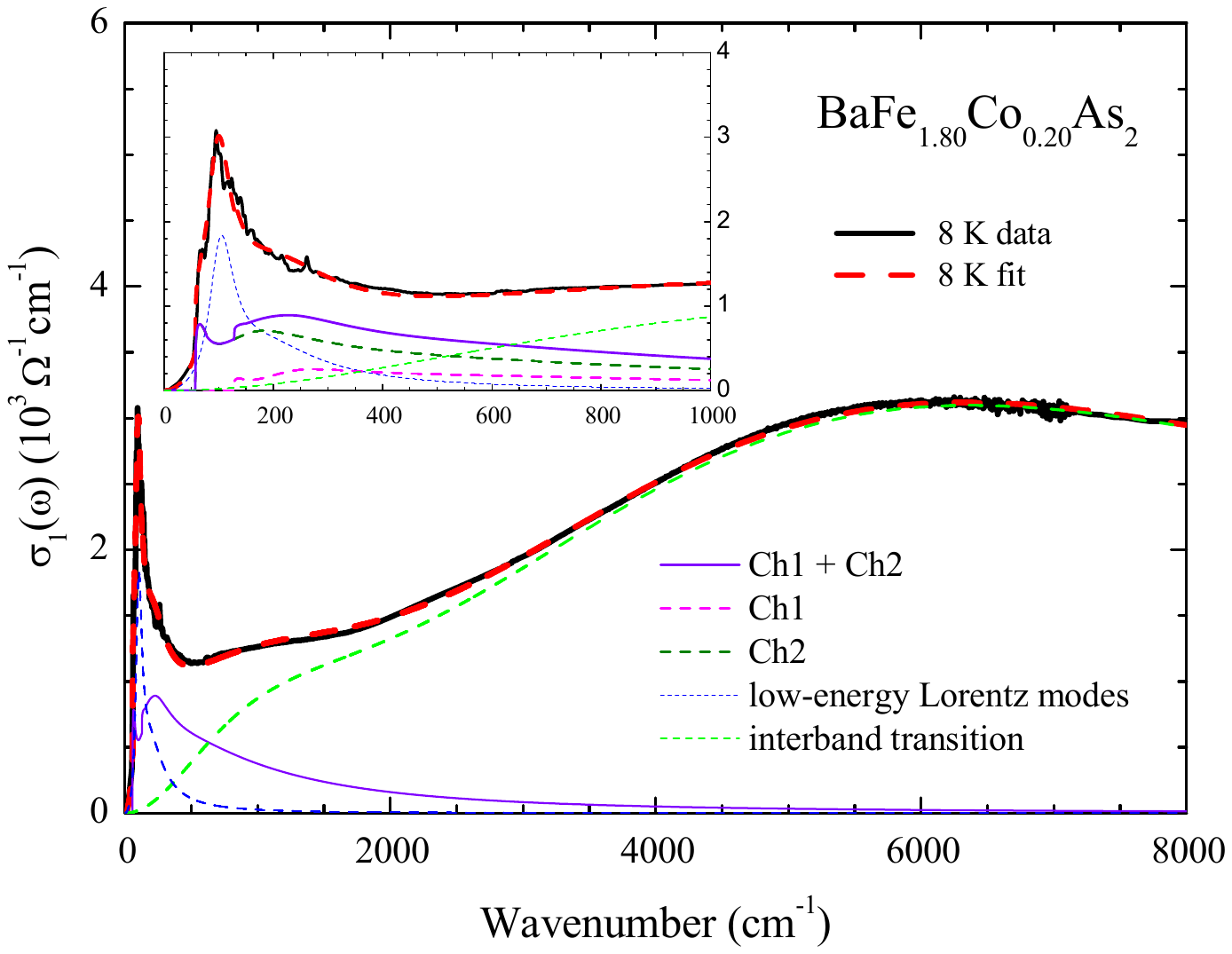}}%
  \vspace*{-1.0 cm}%
\caption{(Color online) Optical conductivity spectrum of BaFe$_{1.80}$Co$_{0.20}$As$_2$ at 8 K (SC state) and its fit in a wide spectral range up to 8000 cm$^{-1}$. In the inset, a magnified view of conductivity is shown to display the two SC channels and the low-energy Lorentz modes, which consists of two Lorentz modes.}
\label{fig5}
\end{figure}

Fig. \ref{fig5} shows data and fit at 8 K (SC state) using the two-parallel-channel approach described above in a wide spectral range up to 8000 cm$^{-1}$. The resulting bosonic spectrum ($I^2B(\omega)$) is shown in Fig. \ref{fig6} (see the blue solid line). Two SC channels (Ch1 and Ch2), the low-energy Lorentz modes, and the interband transition were considered for the fitting. The SC gaps ($\Delta_{Ch1,0}$ and $\Delta_{Ch2,0}$) and the plasma frequencies ($\Omega_{Ch1,p}$ and $\Omega_{Ch2,p}$) for the two SC channels were observed to be 3.6 and 8.0 meV, and 806 and 558 meV, respectively. The same impurity scattering rate ($1/\tau^{op}_{\mathrm{imp}}$) of 15 meV was used for both channels in Eq. (6). The bosonic spectra of the two channels were assumed to be the same as in previous literature \cite{hwang:2016}. The inset in Fig. \ref{fig5} displays a magnified view of the optical conductivities for the two SC channels and the low-energy Lorentz modes located in the low frequency region below 1000 cm$^{-1}$. The low-energy Lorentz modes at the two temperatures (8 and 27 K) were observed to be almost identical. Note that the data could not be properly fitted without including the low-energy Lorentz modes.

\begin{figure}[!htbp]
  \vspace*{-1.0 cm}%
 \centerline{\includegraphics[width=7.0 in]{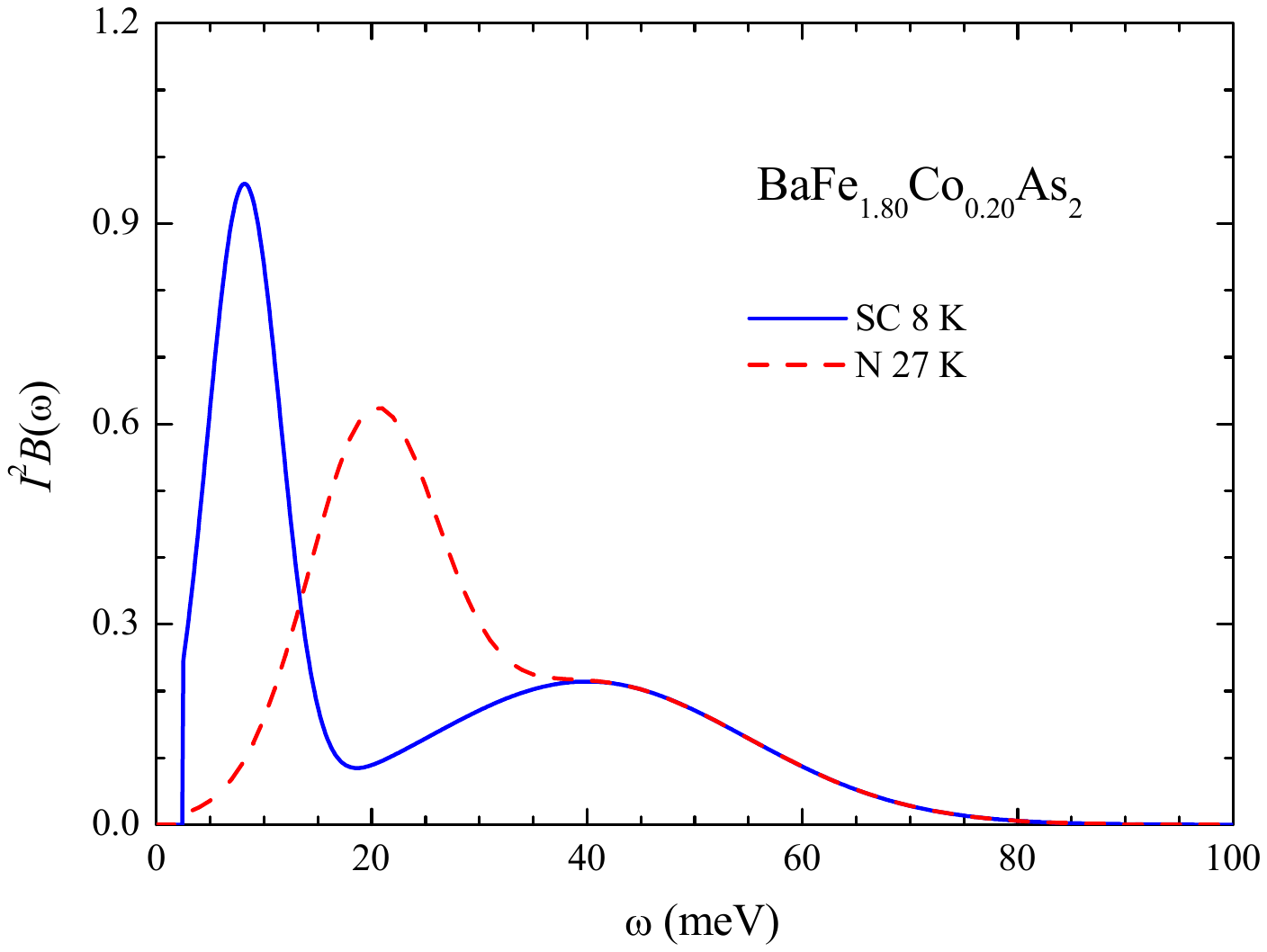}}%
  \vspace*{-1.0 cm}%
\caption{(Color online) Bosonic spectra of BaFe$_{1.80}$Co$_{0.20}$As$_2$ at 8 K (SC state) and 27 K (normal state) obtained using the two-parallel-channel approach and the extended Drude-Lorentz model, respectively.}
\label{fig6}
\end{figure}

Fig. \ref{fig6} shows the resulting $I^2B(\omega)$ at 8 K (SC state) and 27 K (normal state). Note that the resulting $I^2B(\omega)$ shows a similar temperature-dependent trend and a similar spectral width as in previous literature \cite{wu:2010}. Various physical quantities, such as the coupling constant ($\lambda$), the maximum SC transition temperature ($T^{Max}_c$), the SC coherence length ($\xi_{SC}$) and the upper critical field ($H_{c2}$), can be obtained from $I^2B(\omega)$. The coupling constant was obtained from its definition, i.e., $\lambda \equiv 2\int_0^{\omega_c} I^2B(\Omega)/\Omega \: d\Omega$, where $\omega_c$ is a cutoff frequency, 100 meV in our case. The obtained coupling constants were 2.589 and 1.383 at 8 K and 27 K, respectively. $T^{Max}_c$ at 8 K was obtained using the generalized McMillan formula \cite{hwang:2008c}, i.e., $k_B T_c \cong 1.13 \:\: \hbar \omega_{\mathrm{ln}} \:\exp[-(1+\lambda)/(g\lambda)]$, where $k_B$ denotes the Boltzmann constant, $T_c$ denotes the SC transition temperature, $\hbar$ denotes the reduced Planck's constant, $g$ denotes an adjustable parameter between 0 and 1.0, and $\omega_{\mathrm{ln}}$ denotes the logarithmic averaged frequency of $I^2B(\omega)$, which is defined by $\omega_{\mathrm{ln}} \equiv \exp[(2/\lambda) \int_0^{\omega_c} \mathrm{ln}\Omega \: I^2B(\Omega)/\Omega \: d\Omega$]. If $g =$ 1.0, $T_c$ will be its maximum value, $T^{Max}_c$. At 8 K, $\omega_{\mathrm{ln}}$ was found to be 10.41 meV, and $T^{Max}_c$ was found to be 34.1 K, which is larger than the measured value of 22.0 K. Note that the $T_c^{Max}$ is a possible maximum $T_c$, not a real maximum $T_c$. The observation indicates that the $I^2B(\omega)$ is sufficient for explaining superconductivity in the Co-doped Ba122 sample. The time scale of the retarded interaction between electrons via exchanging the mediated bosons is contained in $I^2B(\omega)$ \cite{hwang:2021}. The time scale can be obtained from the first moment of $I^2B(\omega)/\omega$, i.e., $\langle \Omega \rangle \equiv (2/\lambda) \int_0^{\omega_c} \omega' [I^2B(\omega')/\omega'] d\omega'$. The SC coherence length, $\xi_{SC}$, can be obtained from the time scale and the Fermi velocity ($v_{F}$). It can be written in terms of $\langle \Omega \rangle$ and $v_{F}$ as $\xi_{SC} = (1/2) v_F [2 \pi/\langle \Omega \rangle](1/2 \pi) = (1/2)[v_F/\langle \Omega \rangle]$ (or $= (1/2)[\hbar v_F/\hbar \langle \Omega \rangle] $) \cite{hwang:2021}. The estimated $\hbar \langle \Omega \rangle$ was 12.05 meV at 8 K. The reported average Fermi velocity of Co-doped Ba122 systems \cite{richard:2010,brouet:2012}, i.e., $\hbar v_F \cong$ 0.5 eV\AA, was used. The estimated SC coherence length was 20.75 \AA, which is comparable to the reported value of Co-doped Ba122 \cite{kano:2009}. The upper critical field ($H_{c2}$) can also be estimated from the SC coherence length ($\xi_{SC}$) using the Ginzburg–Landau expression, i.e., $H_{c2} = \Phi_0/(2\pi \xi_{SC}^2)$ \cite{kittel:2005}, where $\Phi_0$ denotes the flux quantum. The estimated upper critical field was found to be $\sim$ 75.6 T, which is consistent with the previous results of Co-doped Ba122 \cite{hanisch:2015}. The Ginzburg-Landau parameter ($\kappa$) is defined by the two characteristic length scales ($\lambda_L$ and $\xi_{SC}$) for superconductivity as $\kappa \equiv \lambda_L/\xi_{SC}$ \cite{cyrot:1973}. The estimated $\kappa$ was 99.5, indicating that the Co-doped sample is a type-II superconductor.

The results of the electron-doped BaFe$_{1.80}$Co$_{0.20}$As$_2$ were compared with those of a hole-doped Ba$_{0.49}$K$_{0.51}$Fe$_2$As$_2$. Both samples were similarly overdoped. The SC transition temperatures were 22.0 K and 34.0 K for the electron- and hole-doped Ba122 samples, respectively. The results of the hole-doped sample have been reported previously \cite{lee:2022,lee:2022a}. The energy-dependent distribution of the bosonic spectra of the two systems was observed to be different. The spectral weight of the bosonic spectrum of the Co-doped Ba122 was located in a lower energy region compared with that of the K-doped one (see Fig. \ref{fig6} and Fig. S6 in the Supplementary Materials). This resulted in a larger coupling constant ($\lambda$) of the electron-doped Ba122 (2.59) than that of the hole-doped one (1.68) at 8 K \cite{lee:2022a}. Both the average time scale ($2\pi/\langle \Omega \rangle =$ $2\pi$/(12.05 meV)) and Fermi velocity ($\hbar v_F =$ 0.5 eV\AA) of the bosonic spectrum of the Co-doped Ba122 are larger than those ($2\pi$/(18.72 meV) and 0.38 eV\AA) of the K-doped one, resulting in a longer SC coherence length ($\xi_{SC} =$ 20.75 \AA) for the Co-doped Ba122 compared with that (10.15 \AA) of the K-doped one \cite{lee:2022a}. The $\xi_{SC}$ reflects the size of the Cooper pair, which was larger in the electron-doped Ba122 than that in the hole-doped one. The $\lambda_L$ (206.5 nm) of the Co-doped Ba122 was slightly smaller than that (221.7 nm) of the K-doped one \cite{lee:2022a}. Note that the London penetration depths were obtained using the superfluid plasma frequencies estimated using $\lim_{\omega \rightarrow 0}[-\omega^2\varepsilon_1(\omega)]$. The electron-doped Ba122 exhibits roughly half of the $\kappa$ of the hole-doped one because the former has a twice larger $\xi_{SC}$ but a similar $\lambda_L$ to the hole-doped one. Therefore, the electron-doped Ba122 is a weaker type-II superconductor compared with the hole-doped one. Moreover, the electron-doped Ba122 has the low-energy Lorentz modes, whereas the hole-doped one does not have any such modes (see Figs. \ref{fig4} and \ref{fig5} and Figs. S2 and S3). This difference is due to the doping methods  because electron-doping occurs by replacing the Fe atom in the FeAs plane with a Co atom of a different size from the Fe atom, i.e., the FeAs plane is disordered by Co-doping. However, the FeAs plane, which is known as the charge transport layer, remains intact through hole-doping because the Ba atom in the Ba layer outside the FeAs plane is replaced with a K atom, owing to which K-doped Ba122 is one of the cleanest Fe-based superconductors. Therefore, a finite impurity scattering rate (15 meV) was necessary for fitting the data of electron-doped Ba122, whereas no impurity scattering rate was necessary for fitting the data of hole-doped Ba122 \cite{lee:2022,lee:2022a}. An additional difference between electron- and hole-doped Ba122 is the Lindhard function caused by the different-type carriers, resulting in different effective masses for different Fermi surface sheets \cite{neupane:2011}. We speculate that the differences between the two Ba122 systems describe above may result in different $I^2B(\omega)$. Consequently, the different $I^2B(\omega)$ gives rise to the different SC $T_c$ between the two Ba122 systems.

\section*{Conclusions}

In this study, we investigated an electron-doped (or Co-doped) Ba122 (BaFe$_{1.80}$Co$_{0.20}$As$_2$) using infrared spectroscopy and compared the results with those of a hole-doped (or K-doped) Ba122 (Ba$_{0.49}$K$_{0.51}$Fe$_2$As$_2$). The superfluid plasma frequency ($\omega_{sp}$) of the Co-doped Ba122 was obtained using two independent methods, and the London penetration depth ($\lambda_L$), which are consistent with the reported values of other Co-doped Ba122 systems. The bosonic spectra ($I^2B(\omega)$) at 27 K (normal state) and 8 K (SC state) were obtained using the extended Drude-Lorentz model and the two-parallel-channel approach, respectively. Various physical quantities, such as the coupling constant ($\lambda$), the maximum superconducting transition temperature ($T^{Max}_c$), the SC coherence length ($\xi_{SC}$), and the upper critical field ($H_{c2}$), were obtained from the bosonic spectra. The $T^{Max}_c$ estimated from $I^2B(\omega)$ is larger than the actual $T_c$ measured by the DC transport technique, indicating that the obtained $I^2B(\omega)$ is sufficiently strong to explain the superconductivity in the Co-doped Ba122. Furthermore, the results of the Co-doped Ba122 were compared with the reported results of K-doped one. We expect that our findings will provide insights into the microscopic pairing mechanism of the formation of the Cooper pairs in the doped Ba122 superconductors and useful information for their applications.
\\

\noindent {\bf Acknowledgements} C.K.P. and H.G.L. contributed equally to this work. J.H. acknowledges the financial support from the National Research Foundation of Korea (NRFK Grant Nos. 2020R1A4A4078780 and 2021R1A2C101109811). This research was also supported by BrainLink program funded by the Ministry of Science and ICT through the National Research Foundation of Korea (2022H1D3A3A01077468).
\\
%
% bibliography
%
\bibliographystyle{naturemag}
\bibliography{bib}% Produces the bibliography via BibTeX.

\end{document}